# Compressive Sensing: Performance Comparison Of Sparse Recovery Algorithms

Youness Arjoune and Naima Kaabouch
Electrical Engineering Department
University of North Dakota, Grand Forks, USA

Hassan El Ghazi and Ahmed Tamtaoui
MUSICS Department
INPT, Rabat, Morocco

*Abstract*—Spectrum sensing is an important process in cognitive radio. A number of sensing techniques that have been proposed suffer from high processing time, hardware cost and computational complexity. To address these problems, compressive sensing has been proposed to decrease the processing time and expedite the scanning process of the radio spectrum. Selection of a suitable sparse recovery algorithm is necessary to achieve this goal. A number of sparse recovery algorithms have been proposed. This paper surveys the sparse recovery algorithms, classify them into categories, and compares their performances. For the comparison, we used several metrics such as recovery error, recovery time, covariance, and phase transition diagram. The results show that techniques under Greedy category are faster, techniques of Convex and Relaxation category perform better in term of recovery error, and Bayesian based techniques are observed to have an advantageous balance of small recovery error and a short recovery time.

## I. INTRODUCTION

Next generation communication systems are expected to be smart and fast in sensing the wideband spectrum and identifying the free channels to use [1-2]. Over the last decades, a number of sensing techniques have been proposed, including energy detection [3], Matched filter [4], autocorrelation [5], autocorrelation based Euclidean distance [6], and Bayesian inference method [7]. These techniques are based on a set of Measurements sampled at the Nyquist rate by an Analog/Digital Converter (ADC), which can result in a very high processing time, hardware cost, and computational complexity [8]. In order to overcome these limitations, compressive sensing has been proposed as a solution to decrease the processing time and speed up the spectrum scanning process [9].

Compressive sensing theory asserts that certain signals can be recovered accurately using fewer measurements than the Nyquist/Shannon sampling principle use. As shown in Fig.1, compressive sensing involves three main processes: sparse representation, measurement (encoding), and sparse recovery (decoding) [10].

The first process, sparse representation, consists of representing the signal by a number of projections on a suitable basis. For instance, Wavelet Transform (WT), Fast Fourier Transform (FFT), and Discrete Cosine Transform (DCT) are three examples of sparse representation techniques. A signal $x$ is said to be $K$-sparse if only $K$ elements of its entries are non-zero. Mathematically, this can be written as $\sum \|x\|_1 \leq K$, where $\|.\|_1$ is the $l1$-norm and $K$ is the sparsity level of the signal. If a given signal is not sparse, a simple projection of this signal on a suitable basis can make it sparse.

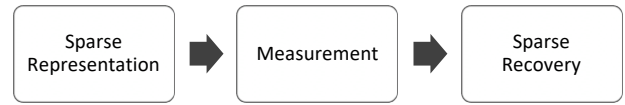

Fig. 1. Compressive sensing processes

The second process, measurement, consists of taking only a few measurements $y \in R^M$ from the sparse signal $x \in R^N$. Mathematically, this can be seen as a multiplication of the sparse signal $x$ by a matrix $\phi \in R^{M \times N}$, where the matrix $\phi$ is the measurement matrix, M is the number of measurements, and N is the length of the sparse signal with M<<N. During this reduction from $R^N$ to $R^M$, compression must preserve the information stored in the $K$-sparse signal necessary to recover the original signal from these few measurements. Measurement matrices can be classified into two categories: random and deterministic matrices. Random matrices present some drawbacks such as costly hardware implementation. Deterministic matrices have been proposed as an alternative to reduce the randomness. Examples of Deterministic matrices include Toeplitz, Circulant matrices [11].

The last process, sparse recovery, aims to recover the sparse signal $x$ from a small set of measurements $y$. Generally, the sparse recovery problem is an underdetermined system of linear equations that needs to be solved using sparse prior. Because this system is underdetermined, the existence and uniqueness of the solution are guaranteed as soon as the signal is sufficiently sparse, and the measurement matrix satisfies the Restricted Isometry Property (RIP) at a certain level [12]. Over the last decades, several sparse recovery algorithms have been proposed [13-24]. These algorithms can be classified into three main categories: Convex and Relaxation, Greedy, and Bayesian category.

Several papers and surveys related to compressive sensing and its application in the context of cognitive radio have been published. For instance, [25] describes a comparison between greedy algorithms. In [26], the authors provided a survey of greedy recovery algorithms. Another survey on compressive sensing and its application was published in [27]. All the aforementioned papers focus either on one of sparse recovery category or on one concept of compressive sensing. The authors in [28] provided a survey on the compressive sensing for cognitive radio. This survey covers all compressive sensing processes: sparse representation, measurement, and sparse recovery algorithms. To the best of our knowledge, a performance comparison between sparse recovery algorithms from all categories has not been published before. Thus, there is a need for detailed review papers that compare and analyze the current sparse recovery algorithms from all categories using different performance metrics. Therefore, in this paper, we propose a classification of these sparse recovery algorithms according to the approach that each one belongs and compare their performances.

The remainder of this article is organized as follows. We first categorize the existing sparse recovery algorithms, then we review a few techniques under each category. The mathematical background of each algorithm is provided. Performance comparison between these algorithms is then investigated, after which concluding remarks are given.

## II. CLASSIFICATION OF SPARSE RECOVERY ALGORITHMS

As shown in Fig. 2, sparse recovery algorithms can be classified into three main categories: Convex and Relaxation, Greedy, and Bayesian category. Techniques under the Convex and Relaxation category solve the sparse signal recovery problem through convex relaxation algorithms. Examples of these techniques include Basis Pursuit [13], Gradient Projection for Sparse Reconstruction (GPSR) [14], and Gradient Descent [15]. Greedy algorithms under the second category recover the sparse signal through an iterative process. Examples of these algorithms include Matching Pursuit (MP), Orthogonal Matching Pursuit (OMP) [16], Compressive Sampling Matching Pursuit (CoSAMP) [17], A*Orthogonal Matching Pursuit (A*OMP) [18], Stagewise Orthogonal Matching Pursuit (StOMP) [19], Generalized Orthogonal Matching Pursuit (GOMP) [20], and Iterative Hard Thresholding (IHT) [21]. The third category, Bayesian framework, solves the sparse recovery problem by taking into account a prior knowledge of the sparse signal distribution. Bayesian techniques can be classified into two types: MAP Estimation and Hierarchical Bayesian framework. MAP Estimation Framework underlines a distribution of the sparse signal $x$ and reconstructs it based on a few measurements. The Hierarchical Bayesian framework introduces one or more variables which control the sparse signal $x$. For instance, Bayesian via Laplace Prior [22], Bayesian via Relevance Vector Machine [23], and Bayesian framework via Belief Propagation [24] are three examples of the hierarchical Bayesian techniques.

## III. METHODOLOGY

As mentioned in the previous section, recovery algorithms can be classified into three categories: Convex and Relaxation, Greedy, and Bayesian category. From each category, we have implemented two algorithms. From the Convex and Relaxation category, we have implemented Basic Pursuit and Gradient Descent algorithms. From the Greedy category, we have implemented Orthogonal Matching Pursuit and Iterative Hard Thresholding algorithms. From the Bayesian category, we have implemented Laplacian prior and Relevance Vector Machine prior algorithms.

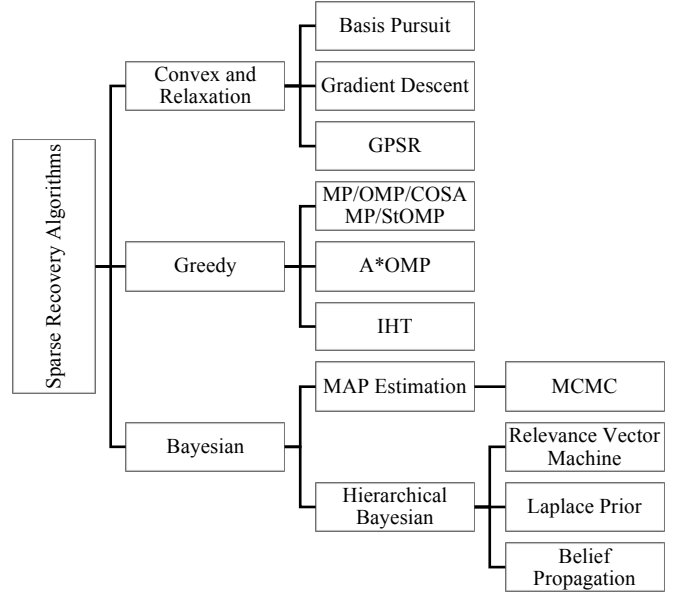

Fig. 2. Classification of sparse recovery algorithms

### A. Convex and Relaxation category
#### 1) Basis Pursuit

Basis Pursuit algorithm finds the sparse vector x with the smallest $l1$-norm that satisfies the equation $\phi x = y$ by using convex optimization, where $y \in R^M$ are the observations, $\phi^{M \times N}$ is the measurement matrix, and $x \in R^N$ is the unknown sparse signal with M<<N [23]. This problem can be formulated as:

$$min\|x\|_1 \quad subject\ to\ \phi x = y \quad (1)$$

Where $\|x\|_1 = \sum_{i=1}^{N} |x_i|$ is the $l1$-norm and $min\|x\|_1$ represent the minimum l1-norm of $x$.
Basis Pursuit is closely connected with linear programming. Therefore, the equation (1) becomes:

$$min\ c^T x \quad subject\ to\ Ax = b\ ;\ x \geq 0 \quad (2)$$

Where $c^T x$ is the objective function , $x \geq 0$ is a set of bounds, $A = (\phi, -\phi)$ , $b = y$, and $c = (1,1)$.

The dual program of the linear program (2) can be written as:

$$\max b^T y \quad subject\ to\ A^T y + z = c\ ;\ x \geq 0 \quad (3)$$

Where z is called the dual variable and $b^T$ is the transpose of b. A fundamental theorem of linear programming states that $(x; y; z)$ is a solution of the linear program (2) if the primal infeasibility $\|b - Ax\|_2$, the dual infeasibility $\|c - z - A^T y\|_2$, and the duality gap $c^T x - b^T y$ are all equal to zeros.

*2) Gradient Descent*

Gradient Descent is an iterative algorithm that finds the sparse solution for the problem (1) where the measurement matrix ϕ satisfies the Restricted Isometric Property (RIP) with an isometric constant $\delta_{2s} < 1/3$. This algorithm calculates iteratively a sparse signal $x \in R^N$ from few measurements $y \in R^M$ using:

$$x = H_s(x + \frac{1}{\gamma}\phi^T r) \quad (4)$$

Where $\gamma = \delta_{2s} + 1/3$, $\phi^T$ is the transpose of the measurement matrix ϕ, r is the residue whose expression is given by $r = y - \phi x$, and $H_s$ is an operator that keeps only the largest magnitude coordinates and sets all other values to zero.

Gradient Descent algorithm starts by initializing the residue $r$ to y, $x$ to zero, and the weight γ to 4/3. Then it updates $x$ for the first iteration using the Eq. 3 and iterates until reaching stopping criteria. More details about this algorithm can be found in [15].

## B. Greedy category
*1) Orthogonal Matching Pursuit*

Orthogonal Matching Pursuit is an iterative Greedy algorithm that computes the best nonlinear approximation of sparse solution for the problem (1). At each iteration, it locates the column k from the measurement matrix ϕ with the largest correlation with the residue $r = y - \phi x$ by taking the higher absolute value of the inner product calculated between each column and the residue using the following formula:

$$k = argmax_j\{|\phi^T r_j|\} \quad (5)$$

Where $\phi^T$ is the transpose of the measurement matrix ϕ. Then, the selected column $k$ is appended to the set $S = S \cup \{k\}$. OMP then estimates the target variable by solving least-squares problem restricted to the columns in S and set all other components of $x$ to zero by using the following formula:

$$(\hat{x}_i)_{/S} = (\phi_{/S})^\dagger . y; (\hat{x}_i)_{/S^c} = 0 \quad (6)$$

Where S is the set of selected columns, $S^C$ denotes complement of the set S, $\phi^\dagger$ denotes the pseudo-inverse of the matrix ϕ, and $(\phi_{/S})^\dagger = (\phi_{/S}^T \phi_{/S})^{-1} \phi_{/S}^T$ is the pseudo-inverse of the matrix ϕ restricted to the set S.

The algorithm updates the residue $r = y - \phi\hat{x}$ and iterates by selecting a new column to be added to the set until the stopping criteria are met [16].

*2) Iterative Hard Thresholding*

Iterative Hard Thresholding algorithm is yet another Greedy algorithm. It finds the sparse signal $x$ subject to the system of linear equation $y = \phi x$ where $y \in R^M$ is an M-dimensional vector of measurements, $x \in R^N$ is the unknown signal to be recovered and $\phi \in R^{M \times N}$ is the measurement matrix. It updates iteratively $x$ using:

$$x^{[n+1]} = H_S(x^{[n]} + \phi^T(y - \phi x^{[n]})) \quad (7)$$

Where Hs is a hard Thresholding operator that sets all the largest elements of $x$ in term of magnitude to zero, $x^{[n]}$ represent the value of $x$ at the iteration n, and $\phi^T$ is the transpose of the measurement matrix ϕ.

Iterative Hard Thresholding algorithm starts by an initialization of the $x^{[0]} = 0$, then at the iteration $n + 1$, it calculates the $x^{[n+1]}$ by calculating $z^n = x^n + \phi^T(y - \phi x^n)$, applying the operator Hs on $z^n$ and iterating until reaching a stopping condition [21].

## C. Bayesian category
*1) Fast Laplace*

Bayesian compressive sensing requires a definition of a joint distribution of the hierarchical model $p(x, \gamma, \beta, y)$. This joint distribution defined as follows in [22]:

$$p(x, \gamma, \beta, y) = p(y/x, \beta) . p(x/\gamma). p(\gamma). p(\beta) \quad (10)$$

Where $\beta = \frac{\sigma^2}{2}$ is the inverse of noise variance, $\gamma$ and $\beta$ are hyperparameters, and the vector of observations $y$ is a Gaussian distribution with zero mean and variance equal to $\beta^{-1}$:

$$p(y/x, \beta) = N(y/\phi x, \beta^{-1}) \quad (11)$$

With a gamma prior placed on $\beta$ as follows:

$$p(\beta/a^\beta, b^\beta) = \Gamma(\beta/a^\beta, b^\beta) \quad (12)$$

The signal model is equivalent to using a Laplace prior on the coefficients $x$:

$$p(x|\gamma) = (\gamma/2)^N \exp(-\gamma/2 \|x\|_1) \quad (13)$$

The Bayesian inference is given by:

$$p(x, \gamma, \beta, \lambda/y) = p(x/y, \gamma, \beta, \lambda) p(\gamma, \beta, \lambda/y) \quad (14)$$

Since $p(x/y, \gamma, \beta, \lambda) \propto p(x, y, \gamma, \beta, \lambda)$, then the distribution $p(x/y, \gamma, \beta, \lambda)$ is a multivariate Gaussian distribution $N(x/\mu, \Sigma)$ with the parameters:

$$\mu = \Sigma \beta \phi^T y, \quad \Sigma = [\beta \phi^T \phi + \Lambda]^{-1}, \text{ and } \Lambda = diag(1/\gamma_i)$$

$p(\gamma, \beta, \lambda/y) = [p(\gamma, \beta, \lambda, y)/p(y)] \propto p(\gamma, \beta, \lambda, y)$ is used to estimate the hyperparameters. We estimate these hyperparameters by maximizing the joint distribution $p(\gamma, \beta, \lambda, y)$ or its logarithm $l$:

$l = \text{Log}(p(\gamma, \beta, \lambda, y)) = -\frac{1}{2}\log|C| - \frac{1}{2}y^T C^{-1} y + N\log(\lambda) - \frac{1}{2}\sum \gamma_i + \frac{\vartheta}{2}\log\left(\frac{\vartheta}{2}\right) - \log\left(\left(\frac{\vartheta}{2}\right)\right) + \left(\frac{\vartheta}{2} - 1\right)\log(\lambda) - \frac{\vartheta}{2}\lambda + \left(a^\beta - 1\right)\log(\beta) - b^\beta \beta$ (15)

The updates of other parameters can be found by solving $\frac{dl}{d\lambda} = 0$ and $\frac{dl}{d\beta} = 0$. The results are given by

$$\lambda = \frac{N - 1 + \frac{\vartheta}{2}}{\sum_i \frac{\gamma_i}{2} + \frac{\vartheta}{2}} \quad (16)$$

$$\beta = \frac{\frac{N}{2} + a^\beta}{\frac{\langle \|y - \phi x\|^2 \rangle}{2} + b^\beta} \quad (17)$$

We can also estimate $\vartheta$ by maximizing (15) with respect to $\vartheta$, which results in solving the equation:

$$\log(\vartheta) + 1 - \psi\left(\frac{\vartheta}{2}\right) + \log(\lambda) - \lambda = 0 \quad (18)$$

Babacan et al. [6] proposed to update only a single element $\gamma_i$ instead of updating the whole vector $\gamma$. This proposition isn't just decreasing the computational requirements, but also is promoting the sparsity. Thus, the logarithm of the joint distribution can be written as:

$l = l(\gamma) = -\frac{1}{2}\left[\log|C_{-i}| - \frac{1}{2}y^T C_{-i}^{-1} y + \frac{\lambda}{2}\sum_{j \neq i}\gamma_j\right] + \frac{1}{2}\left[\log\frac{1}{1+\gamma_i s_i} + \frac{q_i^2 \gamma_i}{1+\gamma_i 2_i} + \lambda \gamma_i\right] = l(\gamma_{-i}) + h(\gamma_i)$

Where $h(\gamma_i) = \frac{1}{2}\left[\log\frac{1}{1+\gamma_i s_i} + \frac{q_i^2 \gamma_i}{1+\gamma_i 2_i} + \lambda \gamma_i\right]$ with $q_i$ and $s_i$ are defined as:

$$s_i = \phi^T_i C_{-i}^{-1} \phi_i \quad (19)$$

$$q_i = \phi^T_i C_{-i}^{-1} y \quad (20)$$

$\frac{dl}{d\gamma_i} = \frac{dh}{d\gamma_i} = 0$ is satisfied at:

$$\gamma_i = \begin{cases} \frac{-s_i(s_i + 2) + s_i\sqrt{(s_i + \lambda)^2 - 4\lambda(s_i - q_i^2 + \lambda)}}{2s_i^2 \lambda}, & \text{if } q_i^2 - s_i > \lambda \\ 0, & \text{otherwise} \end{cases}$$

This algorithm starts by initializing all $\gamma_i$ and $\lambda$ to zero, then tests if $q_i^2 - s_i > \lambda$ and $\gamma_i = 0$. If so, then $\gamma_i$ is added to the model. Otherwise, If $q_i^2 - s_i > \lambda$ and $\gamma_i > 0$, then the algorithm re-estimates $\gamma_i$; otherwise, if $q_i^2 - s_i < \lambda$, then the algorithm prunes I from the model by setting $\gamma_i$ to zero. Finally, the algorithm updates $\Sigma, q_i, s_i,$ and $\lambda$ by using the Eq. 16, and $\vartheta$ using the Eq. 18.

*2) Relevance Vector Machine*

Another probabilistic approach used to estimate the components of $x$ is Relevance Vector Machines (RVM) [23]. This algorithm uses a hierarchical prior to estimate a full posterior on $x$ and on the variance $\sigma^2$, which defines a zero-mean Gaussian prior on each element of $x$. Instead of using the inverse of noise variance, RVM models the prior on $x$ using the precision of a Gaussian density function $\alpha_i$

$$p(x/\alpha) = \prod_{i=1}^{N} \mathcal{N}(x_i / 0, \alpha_i^{-1}) \quad (21)$$

Where $\mathcal{N}(x_i/0, \alpha_i^{-1})$ denotes the Gaussian distribution with a mean equal to zero and a variance $\alpha_i^{-1}$.
In addition, a Gamma prior is considered over $\alpha$ as:

$$p(\alpha/a, b) = \prod_{i=1}^{N} \Gamma(\alpha_i / a, b) \quad (22)$$

Where $\Gamma(\alpha_i / a, b)$ denotes a gamma distribution.
Similarly, a Gamma prior is considered over $\alpha_0 = 1/\sigma^2$:

$$p(\alpha_0/c, d) = \prod_{i=1}^{N} \Gamma(\alpha_0 / c, d) \quad (23)$$

The logarithm of the marginal likelihood can be expressed analytically as:

$\mathcal{L}(\alpha, \alpha_0) = \log(p(y/\alpha, \alpha_0))$
$= \log \int p(y/x, \alpha_0)(p(x/\alpha, \alpha_0)) dx$
$= -\frac{1}{2}[K\log(2\pi) + \log|C| + y^T C^{-1} y]$

Where $C = \sigma^2 I + \phi \Lambda^{-1} \phi^T$; $\Lambda = diag(1/\gamma_i)$.
Thus, the problem of recovering a sparse signal from few measurements in the context of Relevance Vector Machine becomes the search for the hyperparameters $\alpha$ and $\alpha_0$. These hyperparameters are estimated using the EM algorithm:

$$\alpha_i^{new} = \gamma_i / \mu_i \quad (24)$$

Where $\mu_i = 1 - \alpha_i \Sigma_{ii}$, $\Sigma_{ii}$ is the $i^{th}$ diagonal element from $\Sigma$, and $i \in \{1,2,3,\ldots,N\}$.

$$1/\alpha_0^{new} = \frac{\|y - \phi \mu\|_2^2}{k - \sum_i \gamma_i} \quad (25)$$

IV. RESULTS

In this work, we consider a sparse signal of length $N = 1024$ that contains $K$ spikes randomly chosen. The measurement matrix used is a Toeplitz matrix whose size $M \times N$, where M is the number of measurements and N is the length of the sparse signal. Toeplitz matrix reduces the randomness and memory usage and it allows also fast acquisition and recovery. The sparse signal is considered noisy. The generation of the Gaussian noise is done via a random number generator with a standard deviation $\sigma_m = 0.005$. For each sparsity level $K$ or a number of measurements $M$, we consider a Monte Carlo draws by repeating the same experiment 100 times for the same value of $K$ or $M$. In order to compare the performance of sparse recovery algorithms, we used four metrics: recovery error, recovery time, covariance, and phase transition diagram.

Recovery error is a metric to evaluate the error between the original sparse signal and the recovered one. In order to calculate the recovery error, we used the following formula:

$$\text{Err} = \frac{\|x_0 - x_r\|_1}{\|x_0\|_1} \quad (26)$$

Where $x_o$ is the original sparse signal and $x_r$ is the recovered sparse signal.

Recovery time is a metric that measures the time needed by each algorithm to find the solution of the sparse recovery problem.

Covariance is a metric that reflects the correlation between the sensing matrix and the sparse signal. The covariance of $(x_o, x_r)$ is given by:

$$Cov(x_o, x_r) = \text{E}([x_o - \text{E}(x_o)][x_r - \text{E}(x_r)]) \quad (27)$$

Where E is the expectation, $x_o$ is the original sparse signal, and $x_r$ is the recovered sparse signal

Phase transition diagram is a way to determine if a given recovery algorithm can provide good recovery capabilities. It can also be seen as a representation of the success of the sparse recovery algorithm in terms of the probability of success plotted in a phase space of the pair (ρ, δ), where δ=M/N corresponds to the number of samples acquired and ρ=K/M is a ratio of the sparsity level of the signal and the number of samples acquired. For instance, Restricted Isometric Property gives a relationship between the number of measurements, sparsity level, and size of the signal. In order to present these conditions in a clear way, the phase transition diagram can translate these requirements on the signal sparsity level, the size of the signal, and the number of measurements on a plot that separates the phase of success from the failure of recovery of the sparse signal. The explicit expression of the curve ρ (δ; Q) is given by Donoho et al. [29]. They consider the face numbers properties of the projected cross-polytope in order to define the formula of $\rho(\delta; Q)$. This formula is given by:

$$\rho(\delta; Q) = (2\log(1/\delta))^{-1} \text{ as } \delta \to 0 \quad (28)$$

Where δ=M/N, M is the number of measurements, and N is the length of the signal.

In order to measure the asymptotic phase at a point $(\delta, \rho)$, we chose a sequence of $(N, M, K)$ such that δ=M/N and $\rho = K/M$. For each, we generate a Toeplitz matrix A of size $M \times N$ and a random $K$-sparse signal $x$, attempted to recover $x$ from $y = \phi x$ using the six algorithms. In each experiment, we performed Monte Carlo draws to documented the ratio successes to trials as:

$$\hat{\pi} = \frac{\#\ sucesses}{\#\ trials}$$

Fig. 3 shows the recovery error of the six algorithms with respect to the number of measurements with a fixed sparsity level of the signal. When the number of measurement is very low, Basis Pursuit shows better performance than all other techniques. However, when the number of measurements is higher than 130, Bayesian techniques and Orthogonal Matching Pursuit show better performance in term of recovery error which decreases to reach the 0% when the number of measurements is larger than 170.

Fig. 4 shows how the recovery error of these algorithms behaves when the sparsity of a sparse signal of length 1024 is changing, with a fixed number of measurements M=200. According to Fig. 4, as the sparsity of the signal increases, the recovery error of all algorithms increases. Fig. 4 also compares the recovery error of the six sparse recovery algorithms. As can be seen from this figure that Bayesian techniques Orthogonal Matching Pursuit and Basis Pursuit at low sparsity level (k<50) consistently achieve 0% of recovery error. As the sparsity increases, the recovery error of these algorithms increases, but they still show better performance compared to Gradient Descent and Iterative Hard Thresholding techniques.

Fig. 5 shows that the behavior of the Bayesian techniques is opposite to those of Iterative Hard Thresholding and Basis Pursuit techniques. As the number of measurements increases, the recovery time of Bayesian techniques decreases. However, the recovery time for Basis Pursuit and Iterative Hard Thresholding techniques increases as the number of measurements increases.

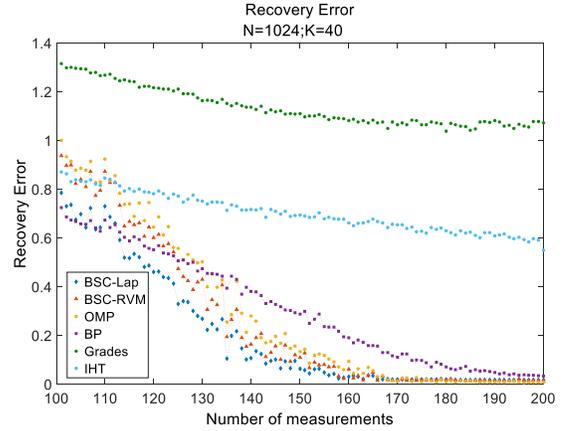
Fig. 3. Recovery error of sparse recovery algorithms with respect to the number of measurements

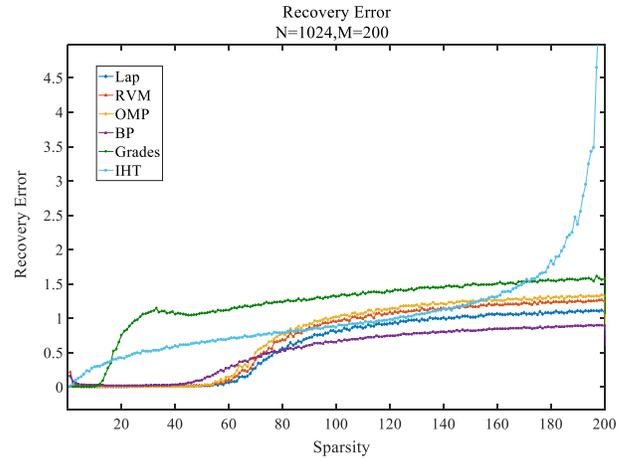
Fig. 4. Recovery error of sparse recovery algorithms with respect to sparsity

Fig. 6 shows that Orthogonal Matching Pursuit and Gradient Descent techniques are faster than all other techniques for all the number of measurements and for all sparsity levels. As shown in this figure, for very low level of sparsity (k<20), Bayesian techniques are faster than all techniques except Orthogonal Matching Pursuit. However, when sparsity is higher than 50, the behavior of the recovery time of Bayesian technique via Laplace Prior increase significantly to become higher than those of the other techniques. However, Bayesian via Relevance Vector Machine technique recovers the signal with an average time below 0.1 sec.

Fig. 7 shows that as the number of measurements increases, the covariance increases reaching 100% when the number of measurements is very high for Bayesian, Basis Pursuit, and Orthogonal Matching Pursuit techniques. On the other hand, Iterative hard Thresholding and Gradient Pursuit show lower correlation.

Fig. 8 shows that the covariance is around 100% when sparsity is between 0 and 40 for Bayesian techniques, Basis Pursuit and Orthogonal Matching Pursuit techniques. The covariance decreases as the sparsity level of the signal increases.

This figure also shows that the Basis Pursuit technique has better performance followed by Bayesian and greedy techniques.

Fig. 9 shows the phase transition diagram of the six sparse recovery algorithms. As can be seen, Basis Pursuit shows best performance followed by Bayesian and greedy techniques.

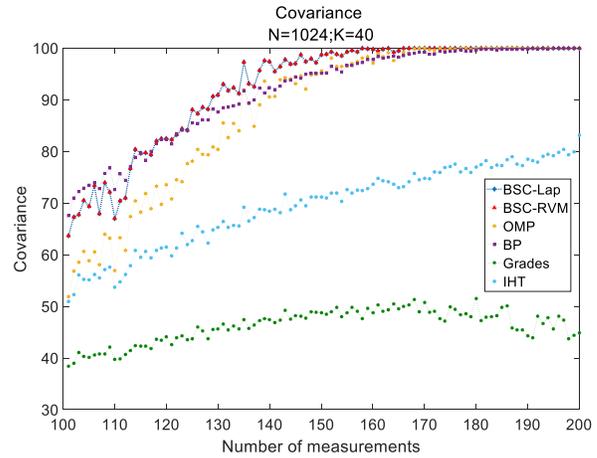

Fig. 7. Covariance of sparse recovery algorithms with respect to the number of measurements

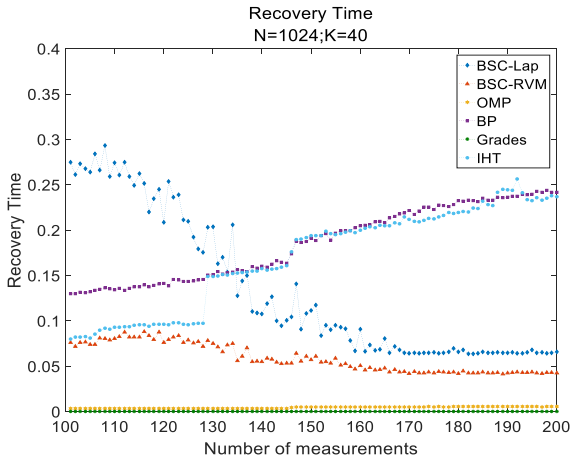

Fig. 5. Recovery time of sparse recovery algorithms with respect to the number of measurements

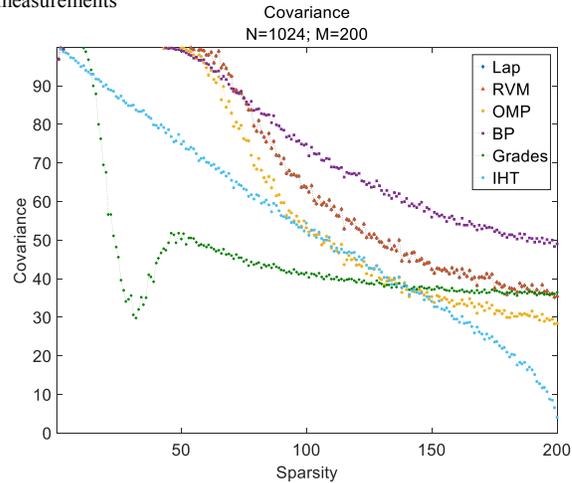

Fig. 8. Covariance of sparse recovery algorithms with respect to the sparsity

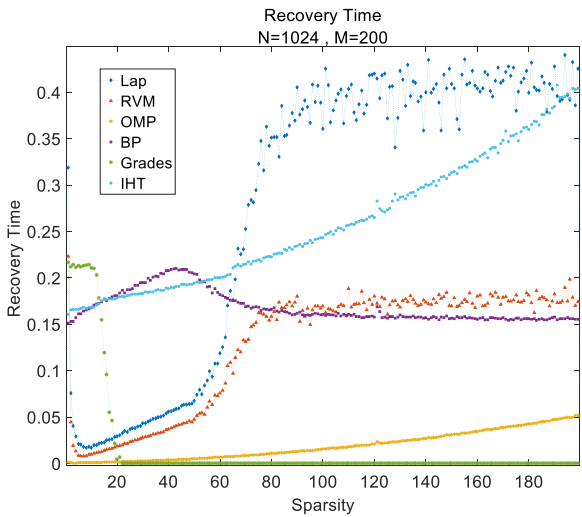

Fig. 6. Recovery time of sparse recovery algorithms with respect to sparsity

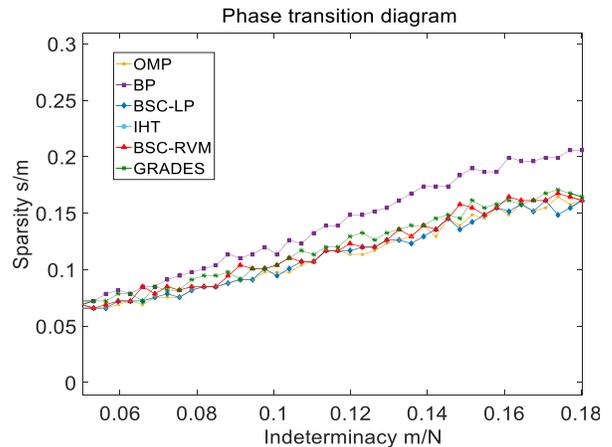

Fig. 9. Phase Transition diagrams of sparse recovery algorithms

Table 1 summarizes the performances of the implemented sparse recovery algorithms. As one can see from this table, Orthogonal Matching Pursuit and Bayesian via relevance vector machine perform better than all other sparse recovery algorithms.

TABLE I. PERFORMANCE COMPARISON OF SPARSE RECOVERY ALGORITHMS

| Sparse Recovery algorithms | Evaluation metrics | | |
|---|---|---|---|
| | *Recovery Error* | *Recovery time* | *Covariance* |
| Basis Pursuit [13] | Small | Slow | High |
| Gradient Descent [15] | High | Fast | Small |
| Orthogonal Matching Pursuit [16] | Small | Fast | High |
| Iterative Hard Thresholding [21] | High | Slow | Small |
| Bayesian via Fast Laplace [23] | Small | Slow | High |
| Bayesian via Relevance Vector Machine [24] | Small | Fast | High |

## V. CONCLUSION

In this paper, we performed a comparison between six sparse recovery algorithms from Convex and Relaxation, Greedy and Bayesian category. The results show that the techniques under Greedy category are faster than the other techniques. However, techniques under Convex and Relaxation category perform better in term of finding the solution to the sparse recovery problem with small errors. Techniques under Bayesian category are a balance between small recovery error and short recovery time. Future work includes the application of the compressive sensing to scanning a wideband spectrum to develop a model for fast data acquisition. This model will be implemented using Universal Software Radio Peripheral (USRP) device and GNU radio Software.